\documentclass[sigconf]{acmart}
\usepackage{xurl}

\copyrightyear{2026}
\acmYear{2026}
\setcopyright{cc}
\setcctype{by}
\acmConference[GLSVLSI '26]{Great Lakes Symposium on VLSI 2026}{June 22--24, 2026}{Canandaigua, NY, USA}
\acmBooktitle{Great Lakes Symposium on VLSI 2026 (GLSVLSI '26), June 22--24, 2026, Canandaigua, NY, USA}
\acmDOI{10.1145/3787109.3815323}
\acmISBN{979-8-4007-2431-2/2026/06}

\usepackage{booktabs}
\usepackage{array}
\usepackage{tabularx}
\usepackage{multirow}
\usepackage{enumitem}

\begin{document}

\title{Orbital AI Computing: Carbon Tradeoffs Across Satellite Scale \vspace{-0pt}}

\author{Nisha Sarwar}
\email{nsarwar@iu.edu}
\affiliation{%
  \institution{Indiana University}
  \city{Bloomington}
  \state{IN}
  \country{USA}
}
\author{Lei Jiang}
\email{jiang60@iu.edu}
\affiliation{%
  \institution{Indiana University}
  \city{Bloomington}
  \state{IN}
  \country{USA}
}  
\author{Fan Chen}
\email{fc7@iu.edu}
\affiliation{%
  \institution{Indiana University}
  \city{Bloomington}
  \state{IN}
  \country{USA}
}

\begin{abstract}
Low Earth Orbit (LEO) computing is emerging for low-latency, globally distributed AI services, enabled by advances in satellite constellations and reusable launch systems.
However, its sustainability remains unclear. Prior work introduces ESpaS, a framework for estimating lifecycle carbon intensity, but models systems using generic datacenter configurations and does not capture modern AI hardware, where power, mass, and compute characteristics vary widely and launch emissions scale with system mass.
In this work, we extend ESpaS with accelerator-aware modeling and evaluate two representative systems: a lightweight Jetson AGX Orin for small satellites and a high-performance DGX H100 enabled by large-payload launch platforms. 
We show that launch emissions act as a fixed carbon overhead: low-mass systems minimize absolute emissions, while high-performance systems amortize this cost more effectively, reducing carbon intensity. Consequently, the space--ground tradeoff is highly sensitive to hardware choice, highlighting the need for accelerator-aware baselines in orbital AI computing. Code available at: \url{https://github.com/nishasarwar/orbital-ai-computing-carbon}.

\vspace{-6pt}
\end{abstract}



\maketitle
\pagestyle{empty}

\vspace{-4pt}
\section{Introduction}
\vspace{-2pt}
Low Earth Orbit (LEO) computing is emerging as a paradigm for low-latency, globally distributed services, enabled by advances in satellite constellations and reusable launch systems. With modern vehicles such as Falcon~9 and Starship reducing launch costs, it is now feasible to deploy compute infrastructure in space, driving interest in \emph{in-orbit computing}, where satellites act as distributed compute nodes for applications from data preprocessing to AI inference. Despite this progress, recent work~\cite{ohs2025dirty} identifies a key sustainability challenge: the carbon footprint of orbital computing is dominated by launch and re-entry emissions, and introduces ESpaS~\cite{ohs2025espas}, a framework and tool for systematically estimating lifecycle carbon intensity across orbital and terrestrial deployments.

However, ESpaS models systems using generic datacenter configurations and does not capture modern AI hardware—an important limitation given that many emerging orbital workloads rely on specialized accelerators with diverse power, mass, and compute characteristics. 
In this work, we focus on \emph{AI-centric orbital systems} and examine how hardware choice influences the space--ground carbon tradeoff. We consider two representative designs: a lightweight Jetson AGX Orin system for small satellites (e.g., CubeSats) and a high-performance DGX H100 system enabled by large-payload launch platforms such as Falcon~9 and Starship, representing opposite ends of the design space. 
We make the following contributions:
\begin{itemize}[leftmargin=*, itemsep=0pt]
\item \textbf{Accelerator-aware modeling.} We extend ESpaS with accelerator-aware hardware profiles reimplemented in Python.
\item \textbf{Cross-scale case study.} We show hardware choice materially affects the space--ground carbon gap across DGX H100 and Jetson AGX Orin.

\end{itemize}

\begin{table}[t]
\caption{Accelerator profile parameters.}
\label{tab:profiles}
\vspace{-8pt}
\centering
\small
\begin{tabular}{p{1.6cm} p{1.7cm} p{2.2cm}}
\toprule
\textbf{Parameter} & \textbf{DGX H100} & \textbf{Jetson AGX Orin} \\
\midrule
Power draw       & 10.2 kW           & 60 W     \\
Peak compute     & 32 FP8 PFLOPS     & 275 INT8 TOPS \\
System mass      & 130.45 kg         & 0.87 kg        \\
\bottomrule
\end{tabular}
\vspace{-0.26in}
\end{table}

\vspace{-0.06in}
\section{Background and Related Work}

\textbf{LEO Computing Systems}.
System design in LEO varies across scales. At the low end, CubeSats—small, low-mass satellites (typically a few kilograms)—are highly constrained in power, thermal capacity, and volume, making them suitable for lightweight edge accelerators such as Jetson-class devices and on-device inference tasks. 
At the high end, modern launch platforms (e.g., Falcon~9, Starship) enable heavier payloads, including large satellites or hosted platforms supporting high-performance accelerators such as DGX-class systems for compute-intensive workloads.
This variation introduces fundamental tradeoffs. Smaller systems minimize absolute launch emissions but offer limited compute capacity, while larger systems incur higher embodied costs yet can amortize these costs over greater compute throughput. As AI workloads increasingly motivate in-orbit computing, these differences in system scale, power, and compute density become central to evaluating sustainability.

\textbf{Related Work}.
ESpaS~\cite{ohs2025espas} is a lifecycle-based tool for estimating the carbon intensity of orbital computing. It decomposes emissions into three components: (1) launch and embodied emissions, (2) operational emissions during the mission, and (3) re-entry emissions. ESpaS applies a unified modeling approach across orbital and terrestrial settings; for terrestrial systems, launch is set to \texttt{None} while retaining a solar- and battery-based configuration.

\textbf{Limitations.}
ESpaS has two key limitations:
(1) The terrestrial baseline lacks a grid-powered pathway and instead models an off-grid system, limiting comparability with modern datacenters that rely on grid electricity and optimized infrastructure.
(2) ESpaS assumes generic datacenter hardware and does not capture modern AI accelerators. This is important because its central finding shows that launch and re-entry emissions dominate the total footprint, implying that embodied emissions scale with system mass and act as a fixed cost. 
In practice, AI systems vary widely in power, mass, and compute density—for example, a DGX H100 node draws $\sim$10.2\,kW at $\sim$130\,kg~\cite{nvidia2024dgx}, while edge devices are orders of magnitude smaller. These differences directly influence how effectively launch emissions are amortized, motivating accelerator-aware modeling for accurate space--ground comparisons.

\vspace{-0.08in}
\section{Methodology}
\subsection{ESpaS extension and validation}
\vspace{-2pt}
We reimplement ESpaS in Python, preserving its original framework and configuration interface (\texttt{SystemConfig}). To ensure fidelity, we reproduce all results reported by~\cite{ohs2025espas}. The reproduced values match the original outputs within numerical precision, confirming correctness and providing a reliable foundation for subsequent extensions.

\vspace{-0.1in}
\subsection{Accelerator-aware modeling}
\vspace{-2pt}
To model modern AI systems, we extend ESpaS with an \textit{accelerator-aware} abstraction. Each system is parameterized by three attributes: (i) peak power draw (kW), determining operational emissions; (ii) peak compute throughput, used for normalization; and (iii) system mass (kg), which influences embodied emissions from launch and manufacturing.This enables consistent cross-scale comparison and captures how mass drives launch emissions as a carbon overhead.

\vspace{-0.1in}
\subsection{Evaluation metrics}
\vspace{-2pt}
We report carbon intensity using three complementary metrics: (i) absolute lifecycle emissions, capturing total environmental cost; (ii) energy-normalized intensity (gCO$_2$e/kWh), reflecting operational efficiency; and (iii) compute-normalized intensity (per TOPS), capturing efficiency per compute unit. DGX throughput is converted from PFLOPS to TOPS (1 PFLOP = 1000 TOPS) for comparability.

\vspace{-0.1in}
\section{Case Study}

\textbf{Impact of hardware choice}.
Tables~\ref{tab:lr-emissions} and~\ref{tab:lifecycle} summarize launch/re-entry emissions and 5-year lifecycle results for DGX H100 and Jetson AGX Orin. A DGX H100 node (130.45\,kg) incurs 22{,}554\,kgCO$_2$e from launch and re-entry on Falcon~9, compared to 150.9\,kgCO$_2$e for Jetson AGX Orin (0.87\,kg), closely reflecting the mass ratio. This confirms that launch emissions scale with system mass and act as a fixed carbon overhead.
Despite higher absolute emissions, DGX H100 achieves lower energy-normalized carbon intensity under orbital deployment (197.77 vs.\ 266.45\,gCO$_2$e/kWh), indicating more effective amortization of launch costs. In contrast, low-power edge systems remain dominated by embodied emissions, resulting in higher carbon intensity per unit energy.

Compute- and energy-normalized metrics provide complementary perspectives. Jetson AGX Orin exhibits slightly lower emissions per unit of compute, whereas DGX H100 achieves lower carbon intensity per unit energy due to improved amortization of fixed launch costs. Nevertheless, orbital systems remain substantially more carbon-intensive than terrestrial baselines, with energy intensity increasing from 34.04 to 161--266\,gCO$_2$e/kWh.

\vspace{-6pt}
\textbf{Key insight}.
Overall, these results demonstrate that the space--ground carbon tradeoff is highly sensitive to hardware characteristics. In particular, system mass, power consumption, and compute density jointly determine how effectively launch emissions are amortized. This underscores the importance of accelerator-aware modeling for sustainable AI in LEO.

\setlength{\tabcolsep}{4pt} 
\begin{table}[t!]
\captionsetup{skip=3pt}
\caption{Launch \& re-entry emissions.}
\label{tab:lr-emissions}
\centering
\small
\begin{tabularx}{\columnwidth}{>{\raggedright\arraybackslash}p{2.0cm} >{\raggedleft\arraybackslash}X >{\raggedleft\arraybackslash}X >{\raggedleft\arraybackslash}X}
\toprule
\textbf{Metric} & \textbf{Earthbound} & \textbf{Falcon 9} & \textbf{Starship} \\
\midrule
\multicolumn{4}{l}{\textit{NVIDIA DGX H100 (130.45 kg)}} \\
\midrule
Total (kgCO$_2$e)      & 0.0 & 22{,}554.5 & 17{,}613.9 \\
Per GPU                & 0.0 & 2{,}819.3  & 2{,}201.7  \\
Per kW TDP             & 0.0 & 2{,}211.2  & 1{,}726.9  \\
Per FP8 PFLOP          & 0.0 & 704.8      & 550.4      \\
\midrule
\multicolumn{4}{l}{\textit{NVIDIA Jetson AGX Orin (0.87 kg)}} \\
\midrule
Total (kgCO$_2$e)      & 0.0 & 150.9      & 117.8      \\
Per kW TDP             & 0.0 & 2{,}514.2  & 1{,}963.5  \\
Per INT8 TOP & 0.0 & 0.55       & 0.43       \\
\bottomrule
\end{tabularx}
\vspace{-0.15in}
\end{table}
\vspace{-0.1in}
\setlength{\tabcolsep}{4pt} 
\begin{table}[t!]
\captionsetup{skip=3pt}
\caption{5-year lifecycle emissions and energy intensity.}
\label{tab:lifecycle}
\centering
\small
\begin{tabularx}{\columnwidth}{>{\raggedright\arraybackslash}p{3.5cm} >{\raggedleft\arraybackslash}X >{\raggedleft\arraybackslash}X >{\raggedleft\arraybackslash}X}
\toprule
\textbf{Metric} & \textbf{Earthbound} & \textbf{Falcon 9} & \textbf{Starship} \\
\midrule
\multicolumn{4}{l}{\textit{NVIDIA DGX H100}} \\
\midrule
Total lifecycle (kgCO$_2$e)       & 15{,}234.9 & 133{,}558.4 & 107{,}639.5 \\
Per TOP (kgCO$_2$e)       & 0.476 & 4.173 & 3.36 \\
Energy intensity (gCO$_2$e/kWh) & 34.04 & 197.77 & 161.91 \\
\midrule
\multicolumn{4}{l}{\textit{NVIDIA Jetson AGX Orin}} \\
\midrule
Total lifecycle (kgCO$_2$e)       & 90.0 & 1{,}003.0 & 803.0 \\
Per TOP  (kgCO$_2$e)                 & 0.33 & 3.65 & 2.92 \\
Energy intensity (gCO$_2$e/kWh) & 34.04 & 266.45 & 215.54 \\
\bottomrule
\end{tabularx}
\vspace{-0.18in}
\end{table}

\section{Limitations and Future Work}
This study has three main limitations:
(1) We evaluate only two accelerator profiles (DGX H100 and Jetson AGX Orin), which, while representative of edge and high-performance systems, do not cover the full spectrum of AI hardware; in particular, TPU-class systems are not included;
(2) Embodied carbon modeling is partial: we include die-level manufacturing (via ESpaS CPA) and launch/re-entry emissions, but omit system-level components (e.g., packaging, chassis, power delivery, assembly), adding upto 20--30\%~\cite{gupta2022chasing};
(3) The die area of the Orin SoC is not publicly available and is approximated as 200\,mm$^2$, introducing uncertainty; we also assume full hardware utilization, which may overestimate real-world efficiency. 
Future work will incorporate additional accelerator classes (e.g., TPUs) and workload-specific utilization. We will also evaluate broader launch vehicles, satellite form factors, and mission durations.

\vspace{-0.1in}

\begin{acks}
\vspace{-2pt}
This work was supported in part by NSF OAC-2417589 and NSF CNS-2143120. 
\end{acks}
\vspace{-0.1in}
\bibliographystyle{ACM-Reference-Format}
\bibliography{carbon}

@article{ohs2025dirty,
  title={Dirty Bits in Low-Earth Orbit: The Carbon Footprint of Launching Computers},
  author={Ohs, Robin and others},
  journal={ACM SIGENERGY Energy Informatics Review},
  year={2025},
}

@misc{nvidia2024dgx,
  title={{NVIDIA DGX H100 Datasheet}},
  howpublished={\url{https://www.nvidia.com/en-us/data-center/dgx-h100/}},
  year={2024}
}

@misc{ohs2025espas,
title={{"ESpaS: Estimator for Space Sustainability"}},
howpublished = {\url{https://zenodo.org/records/15745709}},
year = {2025}
}

@inproceedings{gupta2022chasing,
  title={Chasing carbon: The elusive environmental footprint of computing},
  author={Gupta, Udit and others},
  booktitle={HPCA},
  pages={854--867},
  year={2021},
}

\end{document}